\documentclass[prb,twocolumn,showpacs]{revtex4}
\usepackage{amsmath,bm,graphicx}

\begin{document}

\newcommand{\bild}[5]{
  \begin{figure}
       \resizebox{#2}{!}{\includegraphics{#1}}
       \caption[#4]{#5}
       \label{#3}
  \end{figure}
}

\author{M. Himmerich}
\author{M.~Letz}
\affiliation{Institut f\"ur Physik, Johannes Gutenberg Universit\"at, 
Staudinger Weg~7, D-55099~Mainz, Germany}

\title{The electron gas with a strong pairing interaction:
Three particle correlations and the Thouless instability}

\begin{abstract}
We derive simplified Faddeev type equations for the three particle
T-matrix which are valid in the Hubbard model where only electrons
with opposite spins interact. Using the approximation of dynamical
mean field theory these equations are partially solved numerically for the
attractive Hubbard model. It is
shown that the three particle T-matrix contains a term vanishing $\sim
T^2$ at the Thouless (or BCS) instability where the two--particle
T-matrix diverges.
Based on the three particle term we further derive the low density -- strong
coupling extension for the two-particle vertex function.
We therefore understand our equations as a 
step towards a systematic low density expansion from the weak
coupling BCS theory towards the stronger coupling limit.
\end{abstract}

\date{\today}
\maketitle   

\section{Introduction}
One of the key questions which remains in the understanding of the
unusual superconductors, like the high-T$_c$-materials, some heavy
Fermion compounds or organic superconductors is connected with the
unusual short coherence length of the Cooper pairs. Even without
having clarity on the microscopic pairing mechanism and on the 
evolution of a superconducting phase by e.g. doping an antiferromagnetic 
insulator with holes, this short ranged pair formation points towards 
an effective strong electron--electron interaction. 
It is therefore essential to get a deeper understanding especially of
the thermodynamic properties and of finite temperature quantities of
the electron gas with a strong attractive interaction.
From the 
theoretical understanding the electron gas with an attractive 
interaction has -- besides the atomic one -- only two well understood 
limits. 

The first one is the weak coupling three dimensional (3D) case. 
In this case no two--particle bound state has the ability to 
empty the one--particle continuum and the BCS instability can 
occur for pairs with zero total momentum at the Fermi surface. 
BCS--superconductivity of weakly bound pairs with large 
coherence lengths occurs. These pairs simultaneously
establish pair formation and phase coherence of the 
BCS wave-function. 

The second limit which is well understood is the two--particle 
problem which gives the zero density limit. Solving the 
two--particle problem in one and two dimensions leads for any 
strength of the attractive interaction to the formation of a 
bound state. Since a pair of electrons can be regarded as a boson,
the trapping of a pair in the bound state can be understood as a 
``T=0 condensation'' of a boson into the bound state. Solving the 
two--particle problem in three and more dimensions leads to a clear 
criterion to distinguish between the weak and strong coupling limits. 
In 3D there exists a critical strength of the attractive interaction 
below which no bound state occurs. This marks the weak coupling 
regime where at finite densities the BCS instability will occur. 
Above the critical coupling strength the pair of electrons 
condenses into the bound state. 

The simplest model Hamiltonian where the crossover from the weak
coupling regime towards the stronger coupling limit can be
investigated is the Hubbard model with a non retarded, attractive,
on-site interaction, even 
though it only shows pairing in the s-wave channel. A more complicated
correlation including next neighbor terms can be included later on
where superconducting order parameters can occur which have
non spherical symmetries. Before extending the Hamiltonian we want
to understand the crossover for a simple on-site correlation.

An approach has been widely used to which we refer in agreement 
with many other authors\cite{micnas95,haussmann93} as T-matrix 
approach. It accounts for both of the limits discussed above. 
It exactly solves the two electron problem in the center of mass 
system and therefore gets exact in any dimension and for any 
strength of the interaction at zero density. The non--self-consistent 
T-matrix in the 3D weak coupling case leads to the same equations 
to determine T$_c$ as the BCS theory. The formation of a BCS 
wave-function can be understood as a macroscopic occupation of a
bosonic bound state at the Fermi energy by noninteracting pairs. 
The BCS gap-parameter is then proportional to the density of 
pairs in the condensate wave function. The most problematic part 
in applying the T-matrix to finite densities and to the stronger 
coupling limit is that it neglects pair--pair interactions 
completely in the non--self-consistent formulation\cite{SVR,serene89} 
of the theory. 

Overall, a self-consistent treatment of the two--particle 
T-matrix seems to compare well with quantum Monte Carlo
results.\cite{singer96} However, certain features, like e.g.  
a pseudogap, are either not reported at all from a self-consistent 
treatment of the T-matrix\cite{letz98,keller99} or are 
not seen in the integrated density of states but only in the
k-dependent spectral functions 
at extreme parameter ranges.\cite{kyung98} Therefore,
it is not too surprising that there is a discussion in the 
literature \cite{chen98,beach99} on which degree of 
self-consistency is the right one, which is also motivated 
by the correct emerging of a superconducting phase at low 
temperatures. Using any degree of self-consistency accounts 
in some way
for pair--pair interactions. \cite{micnas95,haussmann93} 
This is however occurring 
in a nonsystematic way, since a small parameter is missing.

In this work we want to tackle the question whether there is a 
systematic way of extending the T-matrix towards the stronger 
coupling and finite density regime while keeping the complete 
dynamics of the equations. As a first step towards this goal 
we investigate the three particle channel. In the three 
particle channel the three body problem is usually decomposed
into two-body fragmentation channels, which define a set of 
Lippmann--Schwinger equations. In general, there are three 
possible two-body fragmentation channels, i.e. three ways 
of combining two out of three particles to interact via a 
two-body force, and a three body interaction. However, for 
the attractive Hubbard model, only two two-body fragmentation
channels are left since electrons with identical spin do not 
interact and an explicit three body interaction is not 
present. Faddeev \cite{faddeev61} derived a set 
of coupled equations for the wave functions in momentum
space, which include the known solution of the 
two--particle problem: the two--particle T-matrix. 
An equivalent formulation was obtained for the 
transition amplitudes in the three body scattering problem.
\cite{alt67} In this work we follow the same spirit and 
calculate the three particle T-matrix as an infinite sum 
over all possible combinations of two--particle T-matrices.
Since we are interested in a dynamic instability, we preserve 
the complete dynamics of the equations in contrast to earlier 
treatments found in the literature\cite{rajaraman67,day67} 
where one was mainly interested in ground state properties. 

In order to be able to do an explicit calculation we use
k-averaged quantities. Doing so, we arrive at equations which 
on one hand become exact in any dimension in the strong 
coupling limit since in the strong coupling limit all 
dispersion disappears. On the other hand, our equations 
meet the dynamical mean--field theory,\cite{metzner89,georges96} 
which gets exact at infinite dimensions for any coupling 
strength since the problem can be mapped in infinite
dimensions onto a local impurity problem because all 
correlations become local. The two--particle T-matrix has 
been studied in the absence of k-dispersion 
\cite{letz98,keller99} and it has been shown that the 
essential physics of the Thouless instability remains
intact. There is however one important difference 
between the limit of infinite dimensions and the strong 
coupling case. In infinite dimensions one clearly has to 
distinguish between zero momentum and momenta which are 
non zero.\cite{keller99} In the strong coupling limit 
the k-dispersion vanishes. Therefore it is not necessary
to treat k=0 separately.\cite{letz98} 

In the two--body problem the instability in the 
particle--particle channel is called `Thouless instability'. 
It is given by the equation
\begin{equation}
\label{eq:thouless}
1 + U \; \chi({\bm{K}} = 0, \Omega = 0) = 0,
\end{equation} 
where $U$ is the attractive on-site interaction and 
$\chi({\bm{K}} , \Omega )$ is the susceptibility in the 
particle--particle channel. We show in this paper that in the three
particle channel  
the equivalent of this equation will be given by a 
determinant of a matrix which becomes zero at the formation
of a three particle bound state. It has been discussed 
earlier that three particle terms should be 
included,\cite{pieri98} although 
they only lead to logarithmic improvement in 2D 
\cite{engelbrecht92,abrikosov57} with respect to a 
low density expansion. However, a systematic low density
expansion should contain the three particle terms as a 
next step beyond the two--particle T-matrix. In the case of
nuclear many body theory the $\alpha$-particle as a four 
particle bound state is of particular low energy and has been 
discussed by R{\"o}pke et al.\cite{roepke98}

The paper is organized as follows: In section \ref{sec1} 
we derive simplified equations for three particles valid in the
Hubbard model where electrons with opposite spin do not interact. The
three particles scatter
in the vicinity of the Fermi surface. 
In section \ref{lowdens} we show how in a close analogy to an earlier
work of Gorkov et al. \cite{gorkov61} the three particle terms can be
used to derive a systematic low density -- strong coupling extension
for the two particle vertex of a pair of electrons with opposite spins.
In section 
\ref{sec:kaverage} we apply a k-average and show that
the equations reproduce the known results in the 
two--particle channel. We pay special attention to keeping
the number of Matsubara frequencies small which are 
needed for the calculation. The analytical extensions to 
correct the sums over Matsubara frequencies are given in
appendices. Having established a numerical solution of 
the two--particle problem which involves only a minimum 
of Matsubara frequencies, we are able to tackle the three 
particle instability in section \ref{sec:three}.

\section{Three particle interactions}
\label{sec1}

All information on the three particle problem is contained in a
six-time correlation function
\begin{equation}
\langle T_{\tau} [
{c_{i,\sigma}}(\tau_1) 
{c_{j,-\sigma}}(\tau_2)
{c_{k,\sigma}}(\tau_3) 
{c_{k',\sigma}^{\dagger}}(\tau_4) 
{c_{j',-\sigma}^{\dagger}}(\tau_5)
{c_{i',\sigma}^{\dagger}}(\tau_6) ] \rangle \;\; .
\end{equation}
Where the $c_{i,\sigma}^{\dagger}, c_{i,\sigma}$ denote creation
and annihilation operators of an electron with spin $\sigma$ on the
lattice site $i$ at imaginary time $\tau$, respectively. The brackets $\langle .. \rangle$
denote the thermal average and
$T_{\tau}$ is the time ordering operator. 
In the first glance it seems to be hopeless to calculate such a
correlation function. However time translational invariance in thermal
equilibrium and spatial translational invariance on the lattice reduce
the number of independent variables to five. Performing all
calculations in frequency and momentum space will further allow via the
Faddeev equations to solve as well in the ''input'' as in the
''output'' channel the occurring two--particle problem by integrating
over the relative momentum and energy of a pair.
We therefore
arrive at functions for the three particle channel which depend on
three dynamical variables only. It is convenient to chose
the total energy and momentum of the triple as one of them.

As already mentioned in the introduction, the three 
particle terms can be built up by a repeated use of 
two--particle interactions.\cite{ethofer69} The special 
simplification in the case of the attractive Hubbard 
model with only a non retarded, on-site attraction, $U<0$, 
is given by the fact that first there is no direct 
three particle interaction and second that the 
correlation $U$ is only effective between two charge 
carriers of different spin which are on the same lattice site. 
Therefore there is only one nontrivial three particle channel where two
charge carriers carry a spin $\sigma$ and the third one has opposite
spin $-\sigma$.
In the following, we use a
convenient short hand notation for the combination of 
summations over Matsubara frequencies and 
${\bm{k}}$-space summations.
\begin{equation}
\label{eq:defom}
\frac{1}{\beta} \sum_1 f(\omega_1) := \frac{1}{\beta} \sum_n
\frac{1}{N} \sum_{{\bm{k}}^1} f(i \omega_n^1,{\bm{k}}^1)
\end{equation}
We use capital Greek letters for Bosonic and small 
ones for Fermionic Matsubara frequencies. Using this 
notation, the Green function of the unperturbed 
system reads
\begin{equation}
{G^0}^{\sigma}(\omega_1) = \frac{1}{i\omega_n^1 - \epsilon({\bm{k}}^1) + \mu},
\end{equation}
where $\mu$ is the chemical potential and $\sigma$ refers 
to the spin index which is for our system of electrons 
(or holes) $\pm 1/2$. The susceptibility in the 
particle--particle channel for a pair of electrons 
with opposite spin is
\begin{eqnarray}
\chi(\Omega_h) &=& \frac{1}{\beta} \sum_1 {G^0}^{\sigma}(\omega_1)
{G^0}^{-\sigma}(\Omega_h - \omega_1) \nonumber \\
&=& \frac{1}{\beta} \sum_n \frac{1}{N} \sum_{{\bm{k}}^1} 
\frac{1}{i\omega_n^1 - \epsilon({\bm{k}}^1) + \mu} \;
\nonumber\\ 
&&\quad\times
\frac{1}{i\Omega_m^h - i\omega_n^1 - \epsilon({\bm{K}}^h - {\bm{k}}^1) + \mu}.
\end{eqnarray}  
The usual two--particle T-matrix results from an infinite 
geometrical series which results from solving the (in this 
case trivially solvable) Bethe--Salpeter equation. 
\begin{equation}
\label{eq:vertex}
\Gamma(\Omega_h) = \frac{U}{1 + U \; \chi(\Omega_h)}
\end{equation}
This is a vertex function which adds all possible 
interactions in the particle--particle channel to 
the usual direct Hubbard interaction. It contains 
the full solution of the two--particle problem in 
the center of mass coordinate system and therefore 
allows the formation of a two--particle bound state 
if the denominator $1 + U \; \chi(\Omega_h)$ becomes
zero. This is occurring first at the chemical 
potential ($i \Omega_n^h=0$) and for zero total 
momentum of the pair ${\bm{K}}^h= {\bm{0}}$ and defines 
the Thouless criterion for the superconducting 
instability.\cite{thouless60} Note that we use a 
different notation for the sign of $\chi(\Omega_h)$ 
then in a previous work \cite{letz98} which is now in
agreement with the book of e.g. Schrieffer 
\cite{schriefferbk} but is different from the one 
used e.g. in Ref. \onlinecite{doniach82}.
For later use we define the functions
\begin{equation}
\label{eq:gamn}
\Gamma^{(l)}(\Omega_h) := \frac{(-U)^l \chi(\Omega_h)}{1+ U \; \chi(\Omega_h)},
\end{equation}
with $l \in \{0,1,2\}$. Here $\Gamma^{(0)}(\Omega_h)$ 
is a two--particle propagator and $\Gamma^{(2)}(\Omega_h)$ 
is the above defined vertex (\ref{eq:vertex}) without 
the single interaction between the particles. The 
definition of the different $\Gamma^{(l)}(\Omega_h)$ 
will be needed later on to be able e.g. to treat the 
contribution of the Hartree term separately.

Following the ideas of Faddeev \cite{faddeev61}, the 
three particle terms can be broken up into a 
two--particle problem plus an additional particle. 
Therefore, we define the following functions.
\begin{equation}
F_1(\omega_1,\omega_2,\omega_{3'}) := -{G^0}^{\sigma}(\omega_1) \; 
{G^0}^{-\sigma}(\omega_2) \; \Gamma(\omega_2 + \omega_{3'})
\end{equation}
This function is illustrated in Fig. \ref{fig:fig2}a. It 
contains all possible repeated correlations between 
particle $(2)$ and $(3')$ and contains the propagator 
for particle $(1)$ which does not interact with any of 
the other two particles. $F$ is already a part of a fully interacting
six-time correlation function. However the time and spatial translational
invariance together with solving the two particle problem in it's
center of mass system has in frequency and momentum space reduced the
function to an expression which contains only three independent
dynamical variables.

The second function which we
define is:
\begin{equation}
F_3(\omega_{3},\omega_{2'},\omega_1) := -{G^0}^{\sigma}(\omega_{3'}) \; 
{G^0}^{-\sigma}(\omega_{2'}) \; \Gamma(\omega_1 + \omega_{2'})
\end{equation}
This function is shown in Fig. \ref{fig:fig2}b. The 
functions $F_1$ and $F_3$ represent the only possible 
two--body fragmentation channels which are present in 
the attractive Hubbard model. A repeated application of 
$F_3$ and $F_1$ turns the frequency $\omega_{2'}$ into
an internal frequency over which a summation has to be 
performed. A diagrammatic illustration of this term is 
shown in Fig. \ref{fig:fig3}.
\begin{equation}
\label{eq:FF}
\frac{1}{\beta} \sum_{2'} F_3(\omega_{3},\omega_{2'},\omega_1) \;
F_1(\omega_1,\omega_2,\omega_{3'}) \; 
\delta_{\omega_2 + \omega_{3'}, \omega_{2'} + \omega_{3}}
\end{equation}
The delta function stems from energy and momentum 
conservation for a pair. It is more convenient to 
introduce a new variable. This is the total energy 
(and momentum) of all three particles $\omega_g =
\omega_1 + \omega_2 + \omega_{3'}$ since it is a 
conserved quantity for the three particle scattering 
events. In this way we define 
\begin{equation}
\label{eq:m}
\begin{align}
M^{(1)}&(\omega_g,\omega_3,\omega_{3'}) := \nonumber\\
&\frac{1}{\beta} \sum_{2'}
F_3(\omega_{3'}, \omega_{2'}, \omega_g -\omega_{2'} -\omega_{3'}) \; \nonumber\\
&\phantom{\frac{1}{\beta}\sum}\times 
F_1(\omega_g - \omega_{2'} -\omega_{3'}, \omega_{2'}+
\omega_{3'}- \omega_3, \omega_3).
\end{align}
\end{equation}
Note that all the Pauli blocking factors which are of 
importance for the discussion of three body interactions 
in nuclear matter \cite{schuck99,beyer99} are included 
in the summation over Matsubara frequencies due to the 
analytic properties of thermal Green functions. 

We can now successively construct all possible interactions 
among three particles by applying alternate scattering in 
the only possible two--body fragmentation channels by a 
repeated application of $M^{(1)}$. For example applying 
$M^{(1)}$ twice gives one additional internal frequency 
(and momentum) and therefore yields an additional
summation.
\begin{equation}
\begin{align}
M^{(2)}&(\omega_g,\omega_3,\omega_{3'}) :=\nonumber\\
&\frac{1}{\beta} \sum_{3''}
M^{(1)}(\omega_g,\omega_3,\omega_{3''}) 
M^{(1)}(\omega_g,\omega_{3''},\omega_{3'}) \nonumber \\
&=: M^{(1)}  \otimes M^{(1)}
\label{eq:prod}
\end{align}
\end{equation}
This defines a formal product for the alternate interaction. 
In appendix \ref{app:metrik} we show that this formal product 
can also be written by introducing a metric.
This product can be repeated several times:
\begin{equation}
\begin{align}
M^{(n)}&(\omega_g,\omega_3,\omega_{3'})=\nonumber\\
&\frac{1}{\beta^{(n-1)}} \sum_{3'',3''',...,3^{(n)}}
M^{(1)}(\omega_g,\omega_3,\omega_{3''})  \;\nonumber\\
&\quad\times
M^{(1)}(\omega_g,\omega_{3''},\omega_{3'''}) 
\; ... \;
M^{(1)}(\omega_g,\omega_{3^{(n)}},\omega_{3'}) \nonumber\\
&=M^{(1)} \otimes M^{(1)} \otimes \; .... \; \otimes M^{(1)}
\end{align}
\end{equation}
In the same spirit one can define a neutral element for the product
defined in Eq. (\ref{eq:prod}).
\begin{equation}
M^{(n)} \otimes E = M^{(n)}
\end{equation}
which leads to:
\begin{equation}
E=\beta \;\delta_{\omega_{3^{(r)}},\omega_{3^{(r')}}} =
\beta \;\delta_{\omega_n^{3^{(r)}},\omega_{n'}^{3^{(r')}}} \;
N \;\delta_{{\bm{k}}_{3^{(r)}}, {\bm{k}}_{3^{(r')}}}   
\end{equation}
It can be seen immediately that the function $E$ fulfills all
properties of a neutral element with respect to the product 
defined in Eq. (\ref{eq:prod}). 
\begin{eqnarray}
M^{(n)} \otimes E &=& \frac{1}{\beta} \sum_{3''}  
M^{(n)}(\omega_g,\omega_3,\omega_{3''}) 
\beta \;\delta_{\omega_{3''},\omega_{3'}} \nonumber \\ &=&
M^{(n)}(\omega_g,\omega_3,\omega_{3'})
\end{eqnarray}
We are now able to express all possible interactions which can occur
between three particles in the attractive Hubbard model. This is given
by the infinite sum over all the three body interaction terms
$M^{(n)}$. It yields a geometrical series which has to be calculated.
\begin{equation}
\begin{align}
M^{\text{tot}}&(\omega_g,\omega_3,\omega_{3'}) = \sum_{n=0}^{\infty}
M^{(n)} (\omega_g,\omega_3,\omega_{3'}) \nonumber\\
&=\beta \; \delta_{\omega_3, \omega_{3'}} + 
M^{(1)} (\omega_g,\omega_3,\omega_{3'}) \nonumber\\
&\phantom{=}+ \frac{1}{\beta} \sum_{3''}
M^{(1)} (\omega_g,\omega_3,\omega_{3''})
M^{(1)} (\omega_g,\omega_{3''},\omega_{3'})
\nonumber\\
&\phantom{=} + \; ...
\end{align}
\end{equation}
which can be formally written as:
\begin{equation}
M^{\text{tot}} (\omega_g,\omega_3,\omega_{3'}) = 
\left [\beta \; \delta_{\omega_3, \omega_{3'}} -
M^{(1)} (\omega_g,\omega_3,\omega_{3'}) \right ]^{-1}
\end{equation}
The inversion has to be such that
\begin{equation}
M^{\text{tot}} \otimes \left [ E - M^{(1)} \right ] = E.
\end{equation}
If on decreasing temperature the function $E-M^{(1)}$ ceases to be
invertible this defines an instability in the three particle
channel. Such an instability marks the appearance of a three particle
bound state. In section \ref{sec:three} we show that in dynamical
mean--field theory, which means in the absence of ${\bm{k}}$-dependence
where $\omega_h \longrightarrow i \omega_n^h$, the condition for the
instability can be expressed as
\begin{equation}
\det [ E_{\text{k-aver}} - M^{(1)}_{\text{k-aver}} ] = 0.
\end{equation}
This determinant has to be calculated for an infinite number of
infinite dimensional arrays. However, restricting the number of
Matsubara frequencies to a finite number allows to calculate the
condition for an instability in the three particle channel.
This is the three particle equivalent of the Thouless instability
(\ref{eq:vertex}) which occurs for scattering of two particles.
We like to mention that a general formulation for an arbitrary
interaction has been given by Ethofer and Schuck.\cite{ethofer69}
Their equations in the p-p-p-channel contain the complete dynamics as our
present work does. Due to the particular simple interaction of the
Hubbard model, our equations obtain the much simpler form given in
this article. 

In nuclear physics \cite{schuck99,beyer99} it is often convenient 
to use simplified formulas for the three--particle T-matrix 
which depend only on one dynamical variable ($z$) instead of 
an expression which depends on three dynamical variables
($\omega_g,\omega_3,\omega_{3'}$). The reason for this is 
that usually only the dependence on the center of mass 
($z \leftrightarrow \omega_g$) is considered. 
Such equations are recovered when all internal initial times are set
to zero when therefore all internal dynamics of the triple is
neglected:
\begin{equation}
M(\omega_g, {\bf k}_3, {\bf k}_{3'}) = 
\lim_{n \longrightarrow \infty \atop
{n'} \longrightarrow \infty}
i \omega_n^3 \; 
i \omega_{n'}^{3'} \; 
M(\omega_g,\omega_3,\omega_{3'})
\end{equation}
In  
section \ref{sec:kaverage} we restrict ourself to an approximation which,
as we will show, preserves the main physics and mathematics 
of the two particle Thouless instability together with the 
complete dynamics of the equations. Afterwards, we are going 
to show how the physics is altered due to an inclusion of 
the three particle terms. 

\section{Systematic low density -- strong coupling extension of the
two particle T-matrix}
\label{lowdens}

In this section we show how the three particle T-matrix can be
incorporated to derive a low density strong coupling extension of the
equations in the particle--particle channel
which lead to the BCS instability. For a
continuous system with a non retarded interaction it has been shown by
Gorkov and Melik-Barkhudaev \cite{gorkov61} that a term which is first
order in low density ($\sim k_F a$) can be used to obtain a low
density correction for the two particle instability. The equations of
\cite{gorkov61} contain however only a correction which is of second
order perturbation theory in the interaction. The term which was
considered in \cite{gorkov61} is illustrated in
Fig. \ref{fig:gorkov}. In the Hubbard model this correction is exactly
zero since a direct interaction between particles with the same spin
is not present. To calculate the next low density correction for the
attractive Hubbard model the first non trivial term is of third order
in the interaction $U$ (the term of second order will only lead to a
Hartree shift). It is illustrated in Fig. \ref{fig:fig3rd}. 
\begin{eqnarray}
\lefteqn{U^0_{\text{eff}}(\Omega,\mu,T) \; \chi(\Omega)= U \; \chi(\Omega) +}
\nonumber \\ && 
\mbox{\hspace*{1cm}}-\frac{U^3}{\beta^3} \sum_{2,3,4} 
G^0(\Omega-\omega_2) G^0(\omega_2) G^0(\omega_3)
G^0(\Omega-\omega_2) \nonumber \\ &&
\mbox{\hspace*{1.5cm}} G^0(\omega_4)
G^0(\Omega+\omega_1+\omega_4-\omega_2-\omega_3) 
\end{eqnarray}
The infinite sum over all the terms up to infinite order of the
interaction $U$ involves the three particle T-matrix
$M^{tot}(\omega_g,\omega_3,\omega_{3'})$ and is shown in
Fig. \ref{fig:lowdens}. To obtain this we define the function:
\begin{eqnarray}
\lefteqn{
H(\Omega-\omega_{3'},\omega_{3'},\omega_{3}) = 
\frac{U}{\beta} \sum_2 G^0(\omega_2) G^0(\Omega-\omega_2)}
\nonumber \\ && 
\mbox{\hspace*{1cm}} \Gamma(\omega_{3'}+\omega_2)
G^0(\omega_2+\omega_{3'}-\omega_3) G^0(\omega_3)
\end{eqnarray}
which can be obtained by replacing one T-matrix in the derivation of
Eq. (\ref{eq:m}) by the bare interaction $U$. Therefore one can obtain the
effective two particle interaction by calculating the product $H
\otimes M^{\text{tot}}$:
\begin{eqnarray}
\label{eq:ueff}
\lefteqn{
U_{\text{eff}}(\Omega,\mu,T) \; \chi(\Omega) = U \; \chi(\Omega)}
\nonumber \\ && \mbox{\hspace*{1cm}} + 
\frac{2}{\beta} \sum_{3'} \left [
H \otimes M^{tot} \right ]
(\Omega+\omega_{3'},\omega_{3'},\omega_{3'}) 
\nonumber \\ && \mbox{\hspace*{1cm}} -
\frac{2U^2}{\beta} \sum_2 G^0(\Omega-\omega_2)
G^0(\omega_2) G^0(\omega_2) \frac{1}{\beta} \sum_3 G^0(\omega_3)
\end{eqnarray}
where the last term is the term of second order which contains the
Hartree term. It has to be subtracted in order to obtain a consistent
inclusion of the Hartree term.
The summation over $\omega_{3'}$ closes all the possible three
particle terms with a hole line making Eq. (\ref{eq:ueff}) a low
density strong coupling correction. 
The factor of $2$ stems from the two possible realizations of the
three particle correction which are $\uparrow \downarrow \uparrow$ and
$\downarrow \uparrow \downarrow$ as combinations for the spins of the
triple. 
The product is given as:
\begin{eqnarray}
\lefteqn{
\left [
H \otimes M^{tot} \right ] (\omega_g,\omega_3,\omega_{3'})}
\nonumber \\ && \mbox{\hspace*{1cm}} = 
\frac{1}{\beta} \sum_4 H(\omega_g,\omega_3,\omega_{4})
M^{tot}(\omega_g,\omega_4,\omega_{3'})
\end{eqnarray}
Equation (\ref{eq:ueff}) leads to the low density -- strong coupling
extension of the two particle instability when inserted into the
Thouless criterion:
\begin{equation}
\label{eq:thouleff}
1 + U_{\text{eff}}(\Omega=0,T,\mu) \; \chi(\Omega=0) = 0
\end{equation}
The correction due to the three particle terms vanishes in the zero
density limit. In the limit of vanishing density the summation over
$1/\beta \sum_{3'} \left [ H \otimes M^{tot} \right ] $ gives
zero and Eq. (\ref{eq:thouleff}) gets identical with the instability
of the two particle T-matrix. The correction due to the product $\left
[ H \otimes M^{tot} \right ] $ vanishes also at high temperatures since
all pair propagators go to zero at infinite temperature
\begin{equation}
\lim_{\beta \rightarrow 0} 
\chi(\Omega) = 0
\end{equation}
Eq. (\ref{eq:thouleff}) gets further exact in the  weak coupling
regime where it reduces for all densities to the Thouless or BCS
condition for a two particle instability since the lowest order
correction to $U$ is of third order in perturbation theory. 

A numerical solution of Eq. (\ref{eq:ueff}) for the limit of
vanishing $k$-dispersion is in progress and will be published
elsewhere \cite{letzhim01}. It is also possible to solve
Eq. (\ref{eq:ueff}) in a selfconsistent way where selfconsistency
has to be achieved for the function $U_{\text{eff}}(\Omega,\mu,T)$. This
might enlarge the applicability of Eq. (\ref{eq:ueff}) even beyond the
low density regime. We further like to mention that the frequency and
momentum dependence of $U_{\text{eff}}(\Omega,\mu,T)$ in Eq. (\ref{eq:ueff})
does not lead to a retarded interaction since the variable $\Omega$
refers to the total momentum and total energy of a pair who's
interaction is being treated.

\renewcommand{\vec}[1]{{\bm{#1}}}

\section{Solution of the two--particle problem}
\label{sec:kaverage}

We need a reliable solution of the two particle problem, if we want
to solve the three--particle problem since the two--particle T-matrix
enters into the Faddeev equations. We will solve the three
particle problem only in a k-averaged approximation, therefore we show in this
section that (i) we are able to reproduce the known results of the
two--particle T-matrix using a k-average and (ii) we demonstrate that
we are able to solve this equations using a minimum of Matsubara
frequencies which is the essential basis to calculate in
Sec. \ref{sec:three} quantities for the three particle problem.

 
In the noninteracting limit $U=0$ the free 
k-averaged Green function is
\begin{equation}
\label{gdef}
G^0(i\omega_n)=\int_{-\infty}^{\infty}
  \frac{D(\epsilon)}{i\omega_n-\omega+\mu}\text{d}\epsilon.
\end{equation}
$D$ is the noninteracting global density of states.
Since we are heading towards a systematic low-density
expansion we are  mainly interested in low band
fillings. Therefore we approximate the density 
of states $D$ by a rectangular shaped one,
which shows the correct analytical behavior
for a 2D system
at the lower and upper band edge. 
\begin{equation}
D(\epsilon)=\frac{1}{W}[\Theta(\epsilon+W/2)-\Theta(\epsilon-W/2)]
\end{equation}
W is the bandwidth. This density of states results in 
\begin{equation}
\begin{align}
\label{eq:gnull}
G^0(i\omega_n)=\frac{1}{W}[&\ln(i\omega_n+W/2+\mu)
  \nonumber\\
&-\ln(i\omega_n-W/2+\mu)].
\end{align}
\end{equation}
The k-averaged pair susceptibility can be calculated from 
the k-averaged Green functions.
\begin{eqnarray}
\chi(i\Omega_n)
&=&\frac{1}{N}\sum_{\vec{K}}\chi(\vec{K},i\Omega_n)\nonumber\\
&=&\frac{1}{\beta N^2}\sum_{\vec{K},\vec{k},m}
      G^0(\vec{k},i\omega_m)
      G^0(\vec{K}-\vec{k},i\Omega_n-i\omega_m)\nonumber\\
\label{chiA}
&=&\frac{1}{\beta}\sum_{m}G^0(i\omega_m)G^0(i\Omega_n-i\omega_m)
\end{eqnarray}
The analytical correction for calculating $\chi$ is shown in 
appendix \ref{app:chi}. We approximate the $\vec{k}$-averaged 
vertex function $\Gamma(i\Omega_n)$ neglecting mean squared 
fluctuations of the pair susceptibility.\cite{letz98}
\begin{equation}
\label{eq:gam}
\Gamma^{(2)}(i\Omega_n)\approx
  \frac{U^2\chi(i\Omega_n)}{1+U\chi(i\Omega_n)}
\end{equation}
$\Gamma^{(2)}(i\Omega_n)$ is used (see Eq. (\ref{eq:gamn})) which
contains the infinite sum of all possible interactions within the pair
except the single interaction. The single interaction $U$ only leads
to a Hartree term when inserted into the selfenergy. The Hartree term
shall be considered separately.
The full interacting Green function is
\begin{equation}
G(i\omega_n)=\left[G^0(i\omega_n)^{-1}-\Sigma^0(i\omega_n)\right]^{-1}
\end{equation}
with the self-energy
\begin{equation}
\label{sigmaA}
\Sigma^0(i\omega_n)=\frac{1}{\beta}\sum_{m=-\infty}^{\infty}
  \Gamma^{(2)}(i\omega_n+i\omega_n)G^0(i\omega_m).
\end{equation}

\subsubsection{Particle number $n$}
\label{sec:tpcn}
Since we need the two particle terms in the three particle T-matrix 
we demonstrate 
in the following that we are able to 
obtain the main results of the two particle T-matrix calculation 
by applying the k-average. We put special 
emphasis on keeping the number of Matsubara frequencies 
small. 

The expectation value of the particle number with a 
certain spin $\sigma$ is calculated from the 
single particle Green function.
\begin{eqnarray}
\langle n_{\sigma}\rangle (\beta,\mu)&=&\lim_{\eta\to 0}\frac{1}{\beta}
  \sum_{l=-\infty}^{\infty}G(i\omega_l)e^{i\omega_l\eta}\nonumber\\
\label{defn}
&=&\frac{2}{\beta}\sum_{l=0}^{\infty}
  \text{Re}G(i\omega_l)+\frac{1}{2}
\end{eqnarray}
The total particle number $n$ is the sum over the two spins.
\begin{equation}
\langle n \rangle =2\langle n_{\sigma} \rangle 
\end{equation}
To obtain Fig. {\ref{graph:n}} we have solved Eqs. (\ref{eq:gnull}) -
(\ref{sigmaA}) non--self-consistently. We have plotted the lines of
constant densities $\langle n_{\uparrow} \rangle 
= \langle n_{\downarrow}\rangle =\langle n \rangle /2$ 
in the plane of the thermodynamic parameters, the
chemical potential without the Hartree contribution 
$\tilde{\mu} = \mu - U \langle n \rangle / 2$ and the temperature T. 
We were able to use only a maximum number of $N_{\text{max}} = 60$
Matsubara frequencies by applying the  
analytical corrections which are shown in appendix \ref{appendix}.
Using these $N_{\text{max}}$ Matsubara frequencies we reach
temperatures which are 
as low as $k_B T = 0.01 [W/2]$. One has to assure
that the range
which is covered by the $N_{\text{max}}$ Matsubara frequencies is
larger than the 
bandwidth of the system
\begin{equation}
2 \frac{(2N_{\text{max}}+1) \pi }{\beta} > W
\end{equation}
In this way less than one hour computer time on a \textsc{pentium} II 
with 450 MHz was needed to obtain figure \ref{graph:n}. All lines 
of constant density collapse into the bound state level on lowering 
the temperature. A result which has been first obtained by 
Schmitt-Rink et al. \cite{SVR} who used
a non conserving approach. In our approach the particle number is
conserved and reaches
1 at the Thouless instability.


\newcommand{\nup}{n_{\uparrow}}
\newcommand{\ndo}{n_{\downarrow}}
\newcommand{\erw}[1]{\left<#1\right>}
\subsubsection{Double occupancy}
\label{sec:tpcnn}
In this section the two particle correlation $\erw{\nup\ndo}$ is computed
from the two particle propagator $\Gamma^{(0)}$. 
\begin{equation}
\label{eq:nn}
\erw{\nup\ndo}=\frac{1}{\beta}\sum_{m=-\infty}^{\infty}
   \Gamma^{(0)}(i\Omega_m)
\end{equation}
In Fig. \ref{graph:nn} we have plotted lines of constant double
occupancy $\langle \nup \ndo \rangle$ in the plane of the
thermodynamical variables $\tilde{\mu}$,T. We have chosen the values
of $\langle \nup \ndo \rangle$ such that some match the square of
$\langle n \rangle$/2 in Fig. \ref{graph:n}. In the limit of high
temperatures or of vanishing interaction ($U \longrightarrow 0$) all
correlations disappear and $\langle \nup \ndo \rangle = (\langle n
\rangle/2)^2 $
To be specific: At high temperatures the line with $\langle \nup \ndo
\rangle = 0.04$ will join the line with $\langle n \rangle = 0.4$ in
Fig. \ref{graph:n} and the one with $\langle \nup \ndo
\rangle = 0.09$ will join the line with $\langle n \rangle = 0.6$.
On lowering the temperature however,
correlations become 
important and $\langle \nup \ndo \rangle$ exceeds the value of  
$(\langle n \rangle/2)^2 $. Close to the Thouless instability $\langle
\nup \ndo \rangle$ even diverges. A diverging value of $\langle \nup
\ndo \rangle$ marks an unphysical behavior since a pair of particles
cannot be more then completely correlated. Therefore $\langle \nup \ndo
\rangle$ should not exceed the value of $\langle n \rangle/2$. The
behavior in the non--self-consistent T-matrix calculation can be
understood since summing over bosonic Matsubara frequencies in
Eq. (\ref{eq:gam}) occupies the bound state with a Bose distribution
and $\langle \nup \ndo \rangle \longrightarrow \infty $ marks the
macroscopic occupation of the bosonic bound state with noninteracting
pairs. A large part of this unphysical behavior is due to the neglect
of pair--pair interactions in the non--self-consistent
T-matrix. Self-consistency seems to repair such a behavior but is uncontrolled
since a small parameter is missing. In order to achieve a systematic
expansion with finally the density as a small parameter we perform a
numerical investigation of
the three particle terms in the next section to answer 
the question, whether this unphysical divergence could be canceled by
the sum of  all three particle scattering processes which contribute 
to $\langle \nup \ndo \rangle$. 


\section{Three particle channel}
\label{sec:three}

In the k-averaged approximation the three particle terms are obtained
by replacing $\omega_k \longrightarrow i \omega_n^k$. Doing so, we
obtain 
\begin{equation}
\label{eq:threekav}
M^{\text{tot}}(i \omega_n^g,i \omega_m,i \omega_{m'}) =
\left [ 
E - M^{(1)}(i \omega_n^g) \right ]^{-1}_{m,m'}
\end{equation}
for the three particle ladder $M^{\text{tot}}$.
This means that in the k-averaged approximation $M^{\text{tot}}$ can be
obtained by inverting the right hand side of Eq. (\ref{eq:threekav})
and multiplying it with a prefactor of $\beta^2$. However, this is an
infinite (due to $i \omega_n^g$) number of inversions of infinite
dimensional matrices which has to be performed. Our efforts described
in the previous section \ref{sec:kaverage} and in appendix
\ref{appendix} enable us to reduce the number of Matsubara frequencies
needed for such an inversion to such a low value that the inversion of
Eq. (\ref{eq:threekav}) can be performed. In the present work we want
to answer the question what is the behavior of the three particle
terms at the two--particle instability.

In Fig. \ref{fig:det} we have plotted the logarithm of the determinant
\begin{equation}
\label{eq:det}
\det \left [ \beta \delta_{m,m'} - M^{(1)}(i \omega_n^g,i
\omega_m,i\omega_{m'}) \right ]
\end{equation}
for one chosen total Matsubara frequency $i\omega_n^g = i\pi /
\beta$, for an attractive interaction $U=-2 [W/2]$ at a chemical
potential $\tilde{\mu} = -0.5[W/2]$ for which $T_c$ 
turns out to be $T_c \approx 0.33301 [W/2]$. On lowering the
temperature the determinant diverges at the Thouless instability
proportional to $(T-T_c)^{-2}$. This is plotted in Fig. \ref{fig:det}
for 60 and 120 as the maximum number of Matsubara
frequencies. In contrast to this, the equivalent of the determinant 
in the particle--particle channel
(see
Eq. (\ref{eq:thouless})) goes to zero linearly in $(T-T_c)$ at the 
instability. 
This has two important consequences. The first one is that even
with a direct inclusion of the three particle terms the double
occupancy will still diverge at the Thouless instability. this is due
to the fact that the double occupancy will be calculated as a sum of
all two particle terms which diverge at the Thouless instability plus
contributions from the three particle terms which vanish at the
two--particle instability and have therefore no chance to cancel the
diverging two--particle terms. The second consequence is therefore
that a correct inclusion of the three particle terms has to go via a
renormalization (possibly even in a selfconsistent way) of the two
particle instability. The renormalization is caused by a third
particle as derived in Eq. (\ref{eq:ueff}).

\section{conclusion}

In this article we present simplified equations for the three
particle terms of the attractive Hubbard model. In particular the
fact that in the Hubbard model electrons with identical spin do not
interact directly with each other, enables us to simplify the equations
considerably. 
Based on the three particle terms we derive in sec. \ref{lowdens}
a low density strong coupling extension of the two particle vertex function
(Eq. (\ref{eq:ueff})). A third particle renormalizes the effective
interaction in the two particle channel and therefore leads to a density
and coupling strength dependent reduction of $T_c$.

In order to solve some of the equations we derived, we applied a
k-averaged approximation which 
becomes formally equivalent to the dynamical mean--field theory where
all correlations become local. Using such a k-averaged approximation
we are able to actually calculate three particle quantities with its
complete dynamics. We
investigate the three particle ladder, where the condition for the
Thouless instability in the two--particle channel (see
Eq. (\ref{eq:thouless})) is now replaced by a determinant of a
matrix. This determinant diverges proportional to $(T-T_c)^{-2}$ as
the Thouless instability is reached. In
the near future it should be possible to extend the present
calculations in three ways. First, the double occupancy as a function
of the thermodynamic parameters including all three particle terms
should be investigated. Calculating this quantity at low densities
should answer the question which degree of self-consistency is the
best one if the two--particle T-matrix is evaluated
self-consistently. This question is currently discussed in the
literature.\cite{chen98,beach99} Second, at temperatures below the
Thouless instability one should continue the determinant (see
Eq. (\ref{eq:det})) to the real axis with respect to $i \omega_n^g$ in
order to obtain a condition for a three particle bound state which
should arise as the determinant vanishes at the real axis. 
The most important extension is a numerical exploration of
Eq. (\ref{eq:ueff}) in the space of the thermodynamic variables $\mu$
and T. This is currently under investigation.

\acknowledgements
M.~L. thanks R.~.J.~Gooding for helpful comments and for reading the
manuscript and P.~Schuck for helpful discussions especially for
pointing towards the importance of ref. \cite{gorkov61}.

\begin{appendix}

\section{Product of three particle interactions involving a metric}
\label{app:metrik}

The formal product defined in Eq. (\ref{eq:prod}) 
can also be written by introducing a metric tensor
\begin{equation}
g^{ij} = \beta \; \delta_{ij} = \beta \;
\delta_{\omega_n^i,\omega_m^j} N \delta_{{\bf k}_i,{\bf k}_j}
\end{equation}
where the last term reminds on the definition in Eq. (\ref{eq:defom}).
This leads to 
a co- and contravariant notation for the matrices using the
Einstein sum convention.
\begin{equation}
M^{(1)}(\omega_g,\omega_3,\omega_{3'}) \longrightarrow M_{33'}
\end{equation}
The formal product then reads:
\begin{equation}
M \otimes N = M_{33''} \; N^{3''}_{\;\;\; 3'} =  
M_{33''} \; g^{3''3'''} \;
N_{3'''3'}
\end{equation}
In the derivation presented in sec. \ref{sec1} and \ref{lowdens}
we did only use the notation defined in
Eq. (\ref{eq:prod}).

\section{Analytical Corrections}
\label{appendix}
The infinite sums over Matsubara frequencies cannot 
be evaluated numerically. 
$N_{\text{max}}$ denotes the maximum number 
of Matsubara frequencies that we want to sum over
numerically. To preserve the features of the infinite sums,
the addends are approximated for large Matsubara frequencies
by terms, which can be summed analytically for an infinite 
number of Matsubara frequencies.

\subsection{The particle number}
\label{app:n}
The real parts of the interacting Green functions in 
(\ref{defn}) are approximated by the noninteracting Green
functions for large Matsubara frequencies (MFs). The noninteracting
Green functions exhibit a $1/(i\omega_n+\mu)$-dependence 
for large MFs.
\begin{equation}
G^0(i\omega_n)\stackrel{\omega_n \mbox{\tiny large}}{\to}
   G_{\delta}(i\omega_n)\equiv\frac{1}{i\omega_n+\mu}
\end{equation}
\begin{eqnarray}
n(\beta,\mu)&\approx&\frac{2}{\beta}\sum_{l=0}^{N_{\text{max}}}
  \text{Re}G(i\omega_l)
  +\frac{2}{\beta}\sum_{l=N_{\text{max}}+1}^{\infty}\text{Re}G_{\delta}(i\omega_l)
  +\frac{1}{2}\nonumber\\
&=&\frac{2}{\beta}\sum_{l=0}^{N_{\text{max}}}\left[\text{Re}G(i\omega_l)
  -\frac{\mu}{\omega_l^2}\right]
  +\frac{\mu\beta}{4}+\frac{1}{2}
\end{eqnarray}
The $1/2$ stems from the sum over the imaginary part where a sum over
$\sin(x)/x$ occurs. \cite{agd}

\subsection{The pair susceptibility}
\label{app:chi}
For the numerical calculation of the pair susceptibility, we look
at the infinite sum (\ref{chiA}). The two Green functions have to be 
approximated by $G_{\delta}$ for values of $i\omega_n$ which are 
far away from the real axis, e.g. for $n>N_{\text{max}}$ and $n<-N_{\text{max}}-1$.
\begin{eqnarray}
\chi_{\delta}(i\Omega_m)&=&\frac{1}{\beta}\sum_{n=-\infty}^{\infty}
   G_{\delta}(i\omega_n)
     G_{\delta}(i\Omega_m-i\omega_n)\nonumber\\
   &=&\frac{\tanh\frac{\beta\mu}{2}}{i\Omega_m+2\mu}
\end{eqnarray}
\begin{eqnarray*}
\Omega_m-\omega_n&=&\frac{2m\pi}{\beta}-\frac{(2n+1)\pi}{\beta}\\
  &=&\frac{[2(m-n)-1]\pi}{\beta}\nonumber\\
  &=&\omega_{(m-n-1)}
\end{eqnarray*}

\begin{eqnarray}
\chi(i\Omega_m)&\approx&\frac{1}{\beta}
   \sum_{n=2N_{\text{max}}-1}^{2N_{\text{max}}}
    \left[G_h(i\omega_n)
       G_h(i\omega_{(m-n-1)})\right.\nonumber\\
     &&\left.-G_{\delta}(i\omega_n)G_{\delta}(i\Omega_m-i\omega_n)
    \right]\nonumber\\
    &&+\chi_{\delta}(i\Omega_m)
\end{eqnarray}
with
\begin{equation}
G_h(i\omega_l)=\left\{\begin{array}{rl} 
      G^0(i\omega_l),&-N_{\text{max}}-1\leq l\leq N_{\text{max}}\\ 
      G_{\delta}(i\omega_l),&\text{otherwise}      
            \end{array}\right.
\end{equation}

In subsequent chapters it will be necessary to approximate
the pair susceptibility for large Bosonic MFs.
Therefore we replace the noninteracting Green functions in 
(\ref{chiA}) by their definitions (\ref{gdef}). This 
results in two integrations over $\epsilon$ and $\epsilon'$.
\begin{eqnarray}
\chi(i\Omega_m)
  &=& \frac{1}{\beta W^2}\sum_{n=-\infty}^{\infty}
   \int_{-\frac{W}{2}}^{\frac{W}{2}}\text{d}\epsilon
   \int_{-\frac{W}{2}}^{\frac{W}{2}}\text{d}\epsilon'
   \frac{1}{i\omega_n-\epsilon+\mu}\nonumber\\
  &&\qquad\times
   \frac{1}{i(\Omega_m-\omega_n)-\epsilon'+\mu}
\end{eqnarray}
 
If we interchange the two integrations with the summation
over $m$, carry out the summation, transform the integration
coordinates to $x=\epsilon+\epsilon'$ and $y=\epsilon-\epsilon'$,
we can integrate over $y$ analytically. The resulting spectral
representation of $\chi$ can be approximated for large 
Matsubara frequencies by $\chi_{W}$.
\begin{eqnarray}
\bar{f_{W}}(\beta,\mu)&=&\frac{-1}{\beta W}
 \ln\frac{1+\cosh[\beta(W/2-\mu)]}{1+\cosh[\beta(W/2+\mu)]}\\
\chi_{W}(i\Omega_m)&=&\frac{\bar{f_{W}}}{i\Omega_m+2\mu}
\end{eqnarray}
Here $\bar{f_{W}}$ denotes the averaged spectral function 
for $\chi$. The function $\chi_W$ has been used to obtain
the results shown in section \ref{sec:tpcn} and \ref{sec:tpcnn}.

\subsection{The self-energy}
Looking at the definition of the self-energy (\ref{sigmaA}) 
we have to approximate the vertex $\Gamma^{(2)}$. It can be 
approximated in first order of $\frac{1}{i\Omega_m}$ for 
large $\Omega_m$ by $U^2\chi_{W}$. Hence we calculate
the following infinite sum.
\begin{eqnarray}
\Sigma_{W}(i\omega_m)&=&
  \frac{1}{\beta}\sum_{n=-\infty}^{\infty}
  U^2\chi_{W}(i\omega_m+i\omega_n)G_{\delta}(i\omega_n)\nonumber\\
&=&-\frac{U^2 \bar{f_{W}}(\beta,\mu)}{i\omega_m+\mu}
  \left[\frac{1}{e^{\beta\mu}-e^{-\beta\mu}}\right]
\end{eqnarray}
The analytically corrected sum over a finite number of 
MFs is a term of the form
\begin{eqnarray}
\Sigma(i\omega_m)&\approx&\frac{1}{\beta}
   \sum_{n=2N_{\text{max}}-2}^{2N_{\text{max}}}
    \left[G^0(i\omega_n)\Gamma^{(2)}(i\omega_m+i\omega_n)\right.\nonumber\\
     &&\left.-G_{\delta}(i\omega_n)U^2\chi_{W}(i\omega_m+i\omega_n)
    \right]\nonumber\\
    &&+\Sigma_{W}(i\omega_m)
\end{eqnarray}
It gives the expression of $\Sigma(i\omega_m)$ which contains the
numerically summed expression for the selfenergy for Matsubara
frequencies close to the real axis and contains the analytical
correction for MFs far away from the real axis.

\subsection{The double occupancy}
The two particle propagator $\Gamma^{(0)}(i\Omega_m)$ 
equals $\chi_{W}(i\Omega_m)$ at large $\Omega_m$. 
Therefore we utilize 
\begin{equation}
\frac{1}{\beta}\sum_{m=-\infty}^{\infty}\chi_W(i\Omega_m)
=\frac{\bar{f_W}(\beta,\mu)}{1-e^{-2\beta\mu}}
\end{equation}
to correct the sum in Eq. (\ref{eq:nn}) analytically. 
 
\subsection{Three particle ladder segment $M^{(1)}$}
For the analytical correction of the sum in Eq. (\ref{eq:FF})
again for the case of vanishing k-dispersion 
we look at the $\omega_{2'}$ depending terms.
Here the terms of the product in Eq. (\ref{eq:FF}) which do not
contain $\omega_{2'}$ are not shown. 
\begin{eqnarray}
&\frac{1}{\beta}\sum\limits_{2'}&
G_{\delta}(i\omega_{2'}+i\omega_{3}-i\omega_{3'})
G_{\delta}(i\omega_g-i\omega_{2'}-i\omega_{3})
\nonumber\\
&&\times G_{\delta}(i\omega_{2'})\left[
-U+U^2\chi_W(i\omega_{3}+i\omega_{2'})
\right]\\
&=&\frac{
-U\left[
\bar{a}^2 \bar{b}bb'+
U\left(
bb'+a_0 o+a_0^2 bb'
\right)\bar{f_W}(\beta,\mu)
\right]
}{
a\bar{a}\bar{b}bb'cc'
}\nonumber
\end{eqnarray}
with
\begin{eqnarray}
a_0&=&e^{\beta\mu}\nonumber\\
a&=&e^{\beta\mu}+1\nonumber\\
\bar{a}&=&e^{\beta\mu}-1\nonumber\\
\bar{b}&=&3\mu+i\omega_g\nonumber\\
b&=&\mu+i\omega_3\\
b'&=&\mu+i\omega_{3'}\nonumber\\
c&=&2\mu-i\omega_3+i\omega_g\nonumber\\
c'&=&2\mu-i\omega_{3'}+i\omega_g\nonumber\\
d&=&i\omega_3+i\omega_{3'}\nonumber\\
o&=&\omega_g^2+\omega_3\omega_{3'}-i\omega_g (d-4\mu)-4d\mu+2\mu^2\nonumber
\end{eqnarray}
\end{appendix}
This allows to calculate the three particle function
$M(\omega_g,\omega_3,\omega_{3'})$ for the case of a neglected
k-dispersion. In section \ref{sec:three} the determinant of $E-M$ close
to the Thouless instability is investigated.


\eject

\begin{figure}
\centerline{\rotatebox{0}{\resizebox{8cm}{!}{\includegraphics{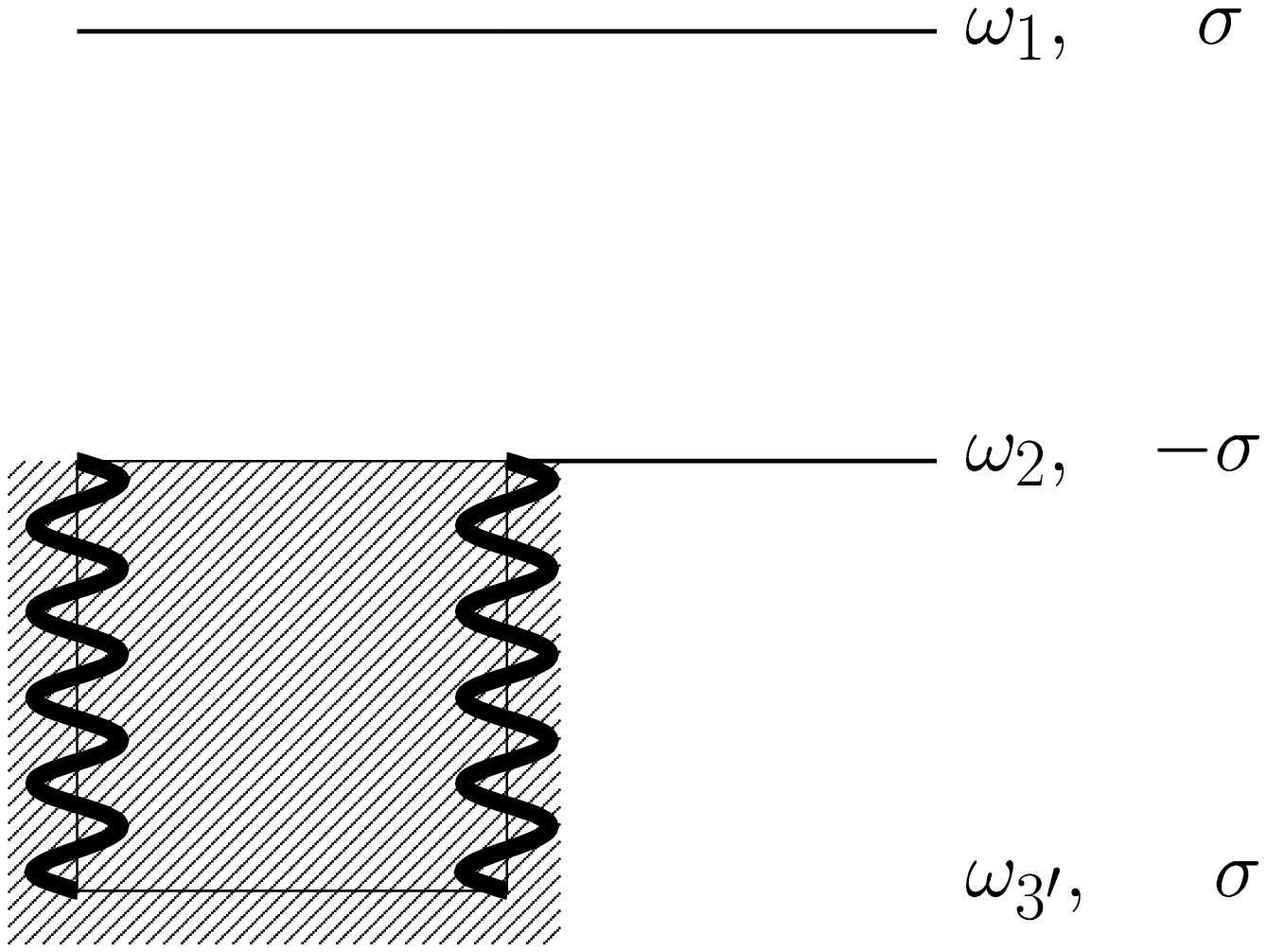}}}}
\centerline{\rotatebox{0}{\resizebox{8cm}{!}{\includegraphics{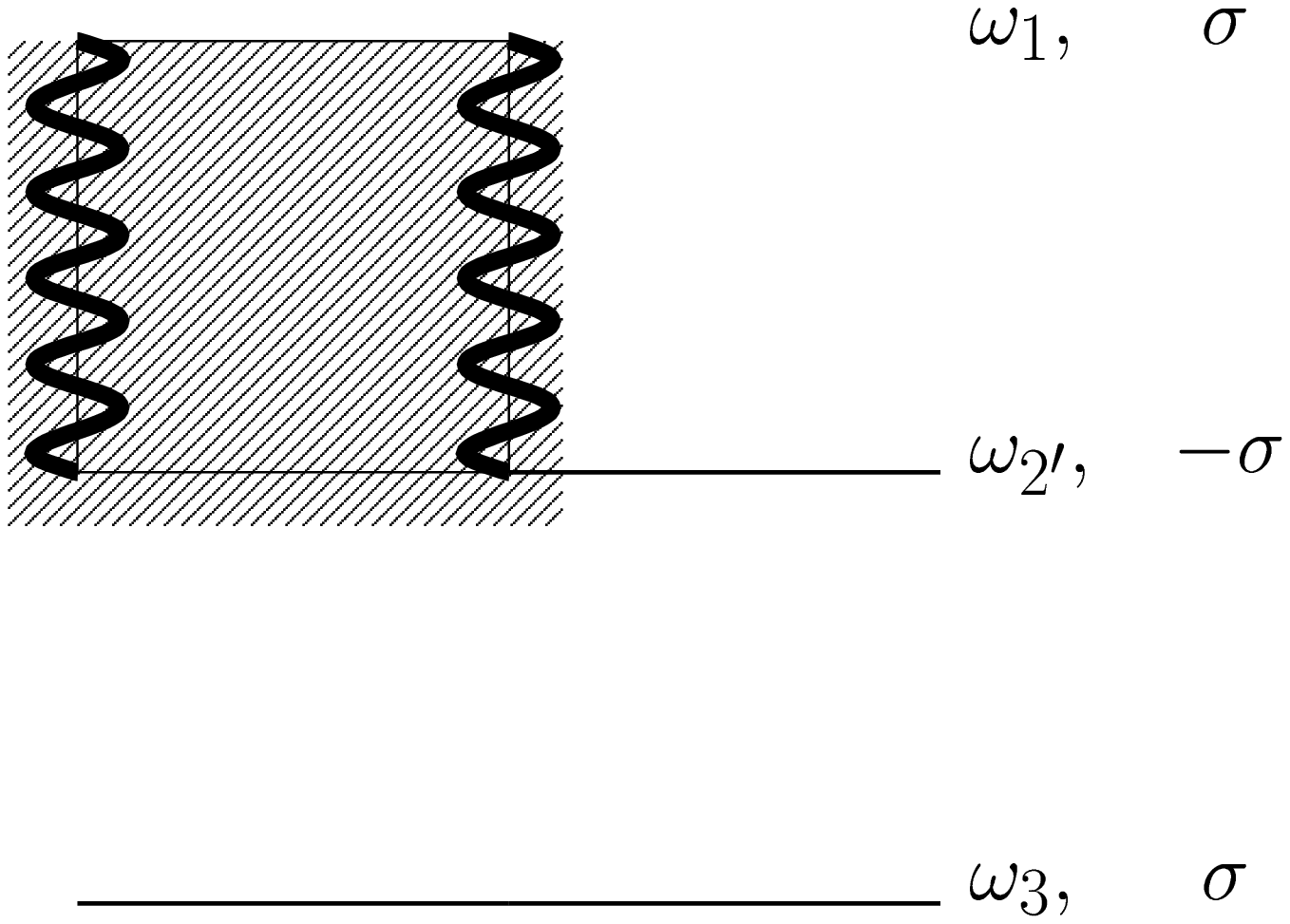}}}}
\caption{The function $F_1(\omega_1,\omega_2,\omega_{3'})$ is shown
diagrammatically in Fig. \protect{\ref{fig:fig2}}a . The shaded area
denotes the  T-matrix and represents all the possible 
interactions between particle $2$ and $3$. Fig. \protect{\ref{fig:fig2}}b
shows the function $F_3(\omega_{3},\omega_{2'},\omega_1)$.
}
\label{fig:fig2} 
\end{figure}

\begin{figure}
\centerline{\rotatebox{0}{\resizebox{12cm}{!}{\includegraphics{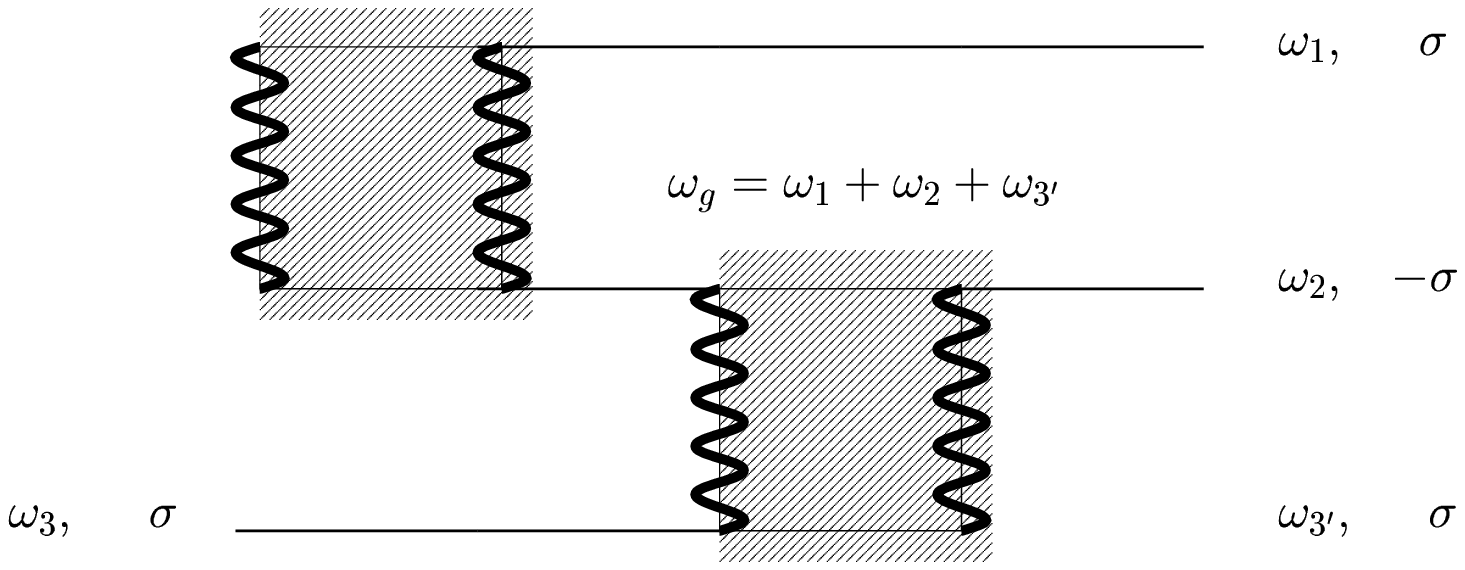}}}}
\caption{The repeated application of $F_1$ and $F_3$ gives the
function $M^{(1)}(\omega_g,\omega_3,\omega_{3'})$. This is shown as a
diagrammatic representation.
}
\label{fig:fig3}
\end{figure}

\begin{figure}
\centerline{\rotatebox{0}{\resizebox{10cm}{!}{\includegraphics{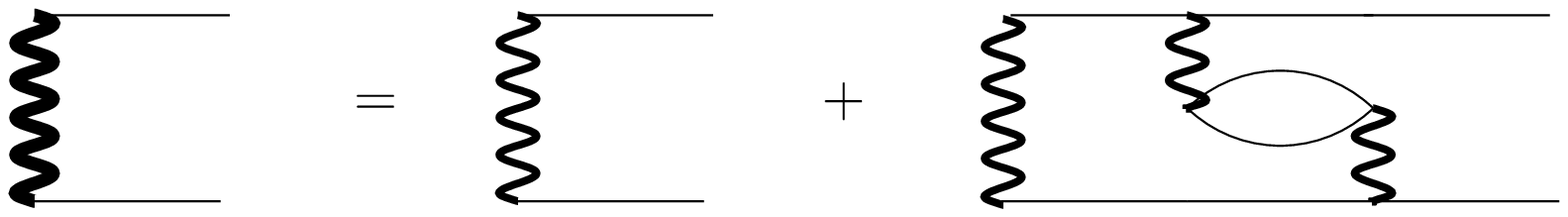}}}}
\caption{Low density correction which was considered in 
\protect{\cite{gorkov61}}. The last term does not give any contribution for the
attractive Hubbard model since a direct interaction between particles of
identical spin is not present.
}
\label{fig:gorkov}
\end{figure}

\begin{figure}
\centerline{\rotatebox{0}{\resizebox{10cm}{!}{\includegraphics{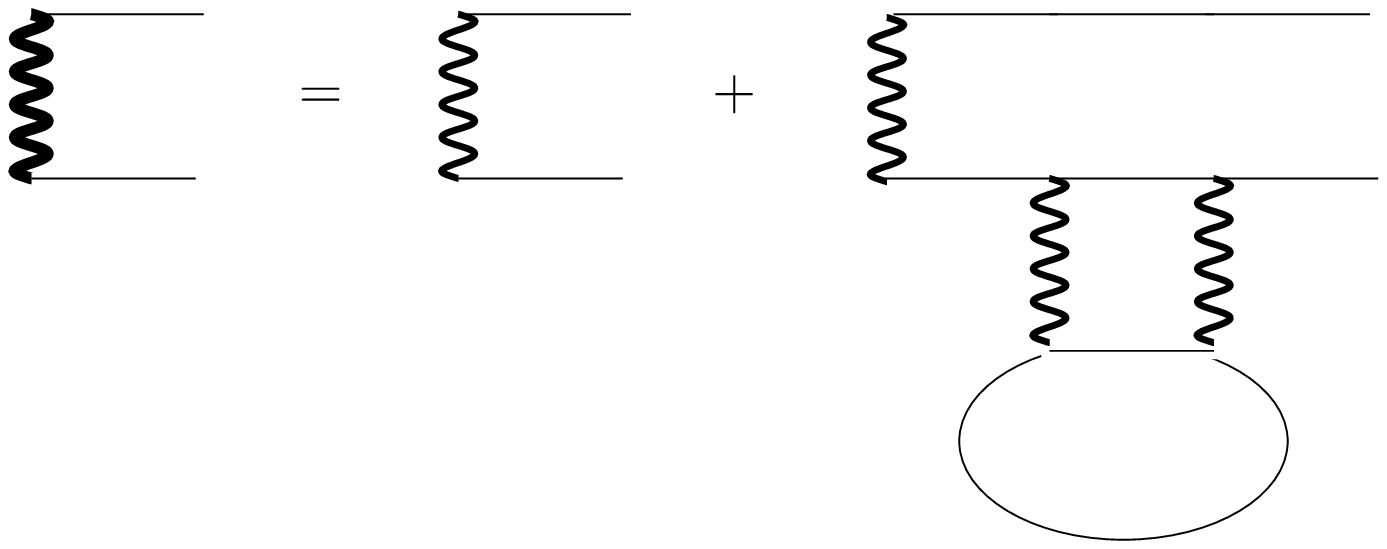}}}}
\caption{First non--vanishing term for a systematic low density
correction which is of third order in the interaction $U$.
}
\label{fig:fig3rd}
\end{figure}

\begin{figure}
\centerline{\rotatebox{0}{\resizebox{10cm}{!}{\includegraphics{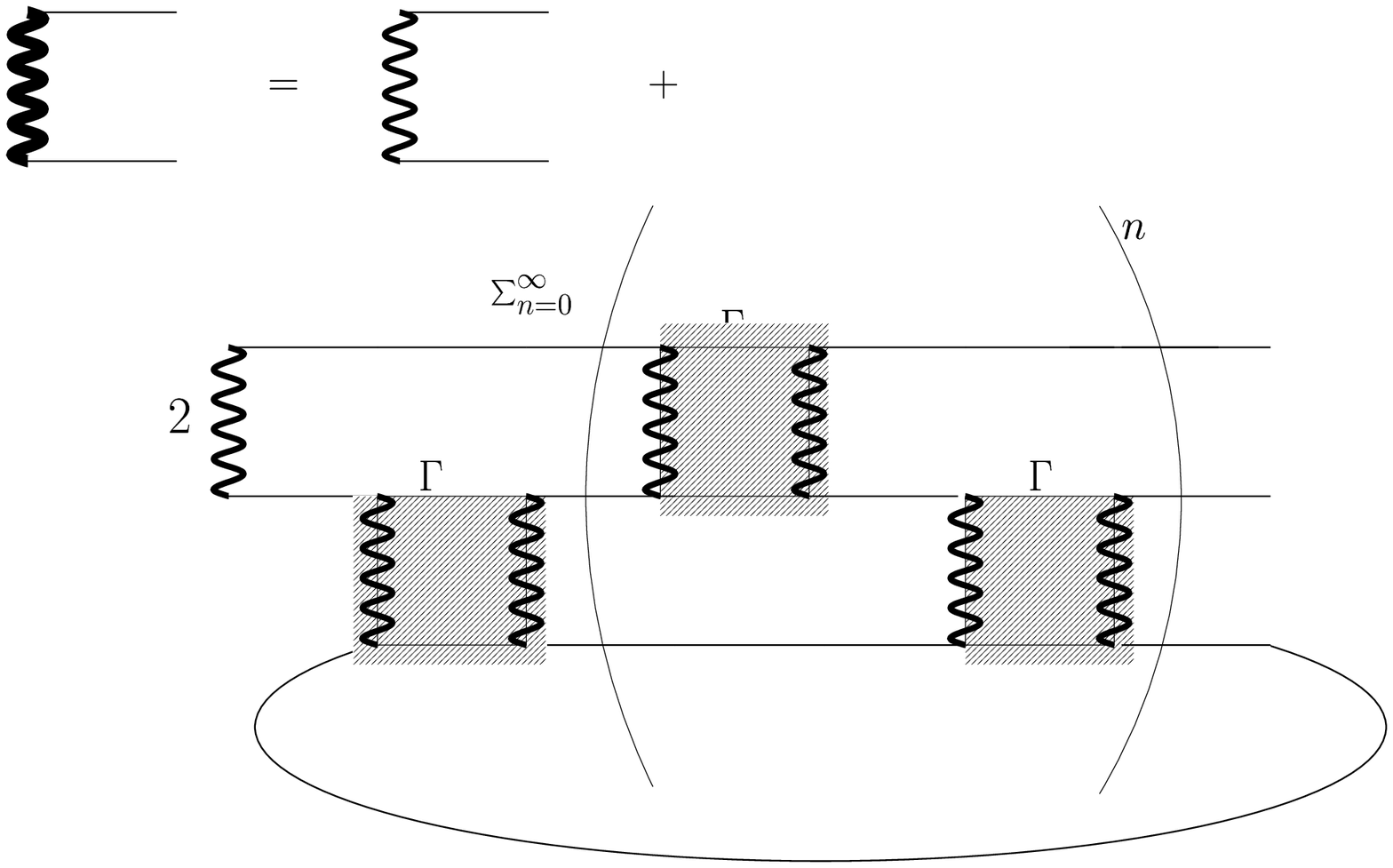}}}}
\caption{Infinite sum over all orders of the interaction $U$ which
gives the complete low-density strong coupling extension formulated in
Eq. (\protect{\ref{eq:ueff}}) with the exception of the Hartree term
which has to be subtracted.
}
\label{fig:lowdens}
\end{figure}

\begin{figure}
\label{graph:n}
\centerline{\rotatebox{-90}{\resizebox{8cm}{!}{\includegraphics{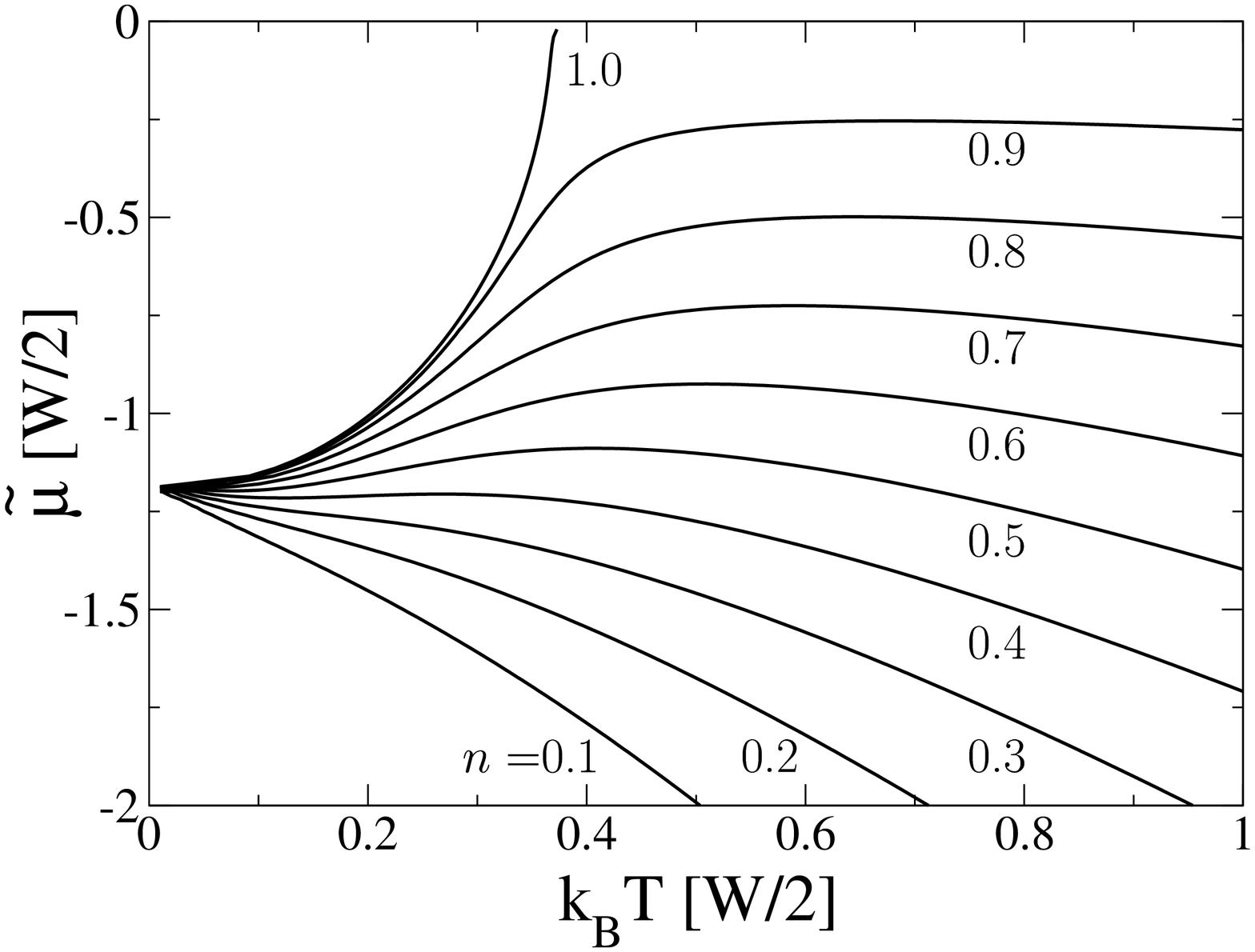}}}}
\caption{Lines of fixed
particle numbers on the $T$-$\tilde{\mu}$-plane. The attractive interaction
$U$ is equal to the bare bandwidth $W$ and $n=1$ denote half filling. At low
temperatures all lines of constant density collapse into the
two--particle bound state.\vspace*{1cm}}
\end{figure}

\begin{figure}
\label{graph:nn}
\centerline{\rotatebox{0}{\resizebox{9cm}{!}{\includegraphics{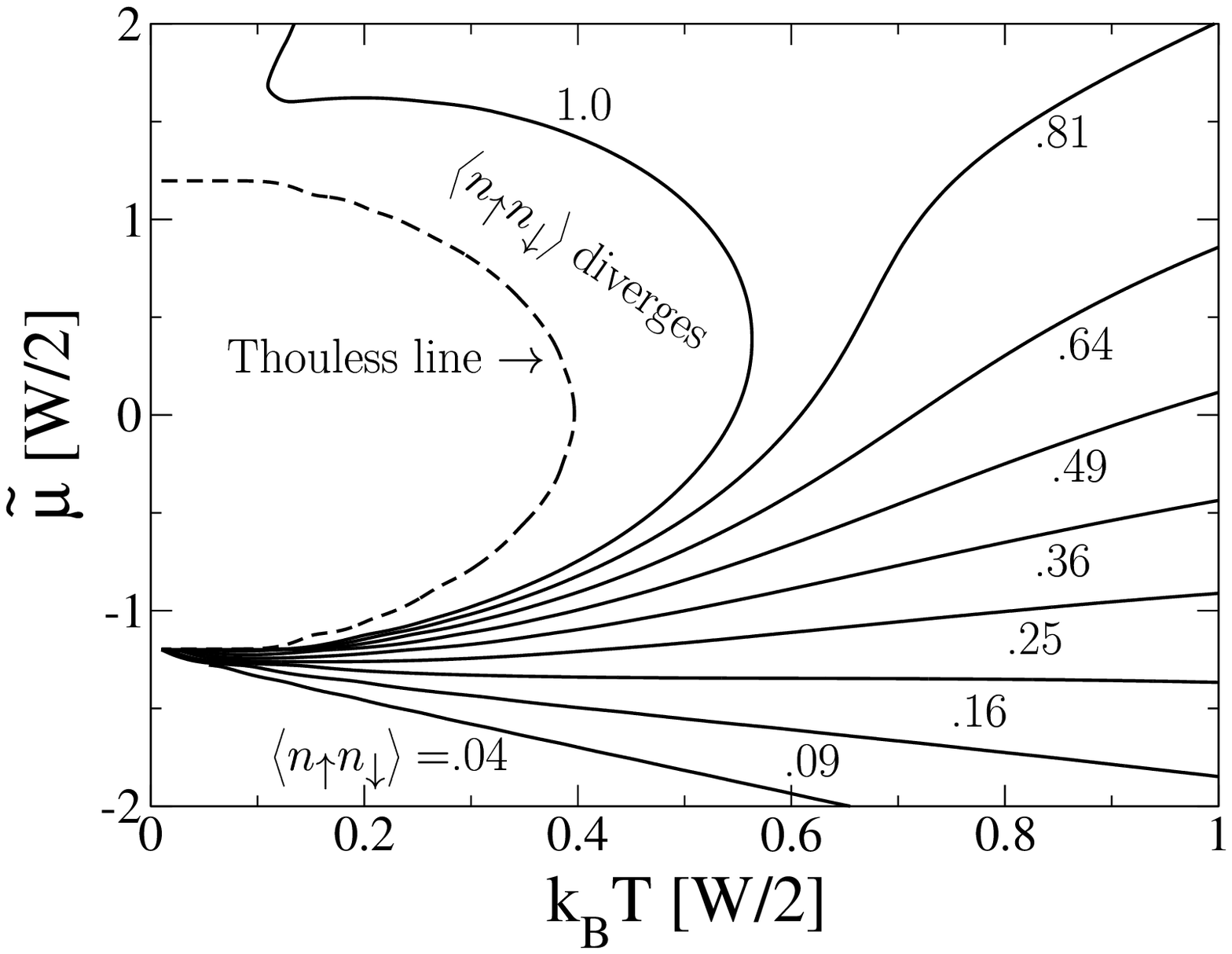}}}}
\caption{Lines of fixed
$\erw{\nup\ndo}$ on the $T$-$\tilde{\mu}$-plane. At the Thouless instability 
$\langle \nup\ndo \rangle $ divergences. The attractive interaction
$U$ is equal to the bare bandwidth $W$. 
}
\end{figure}


\begin{figure}
\centerline{\rotatebox{0}{\resizebox{10cm}{!}{\includegraphics{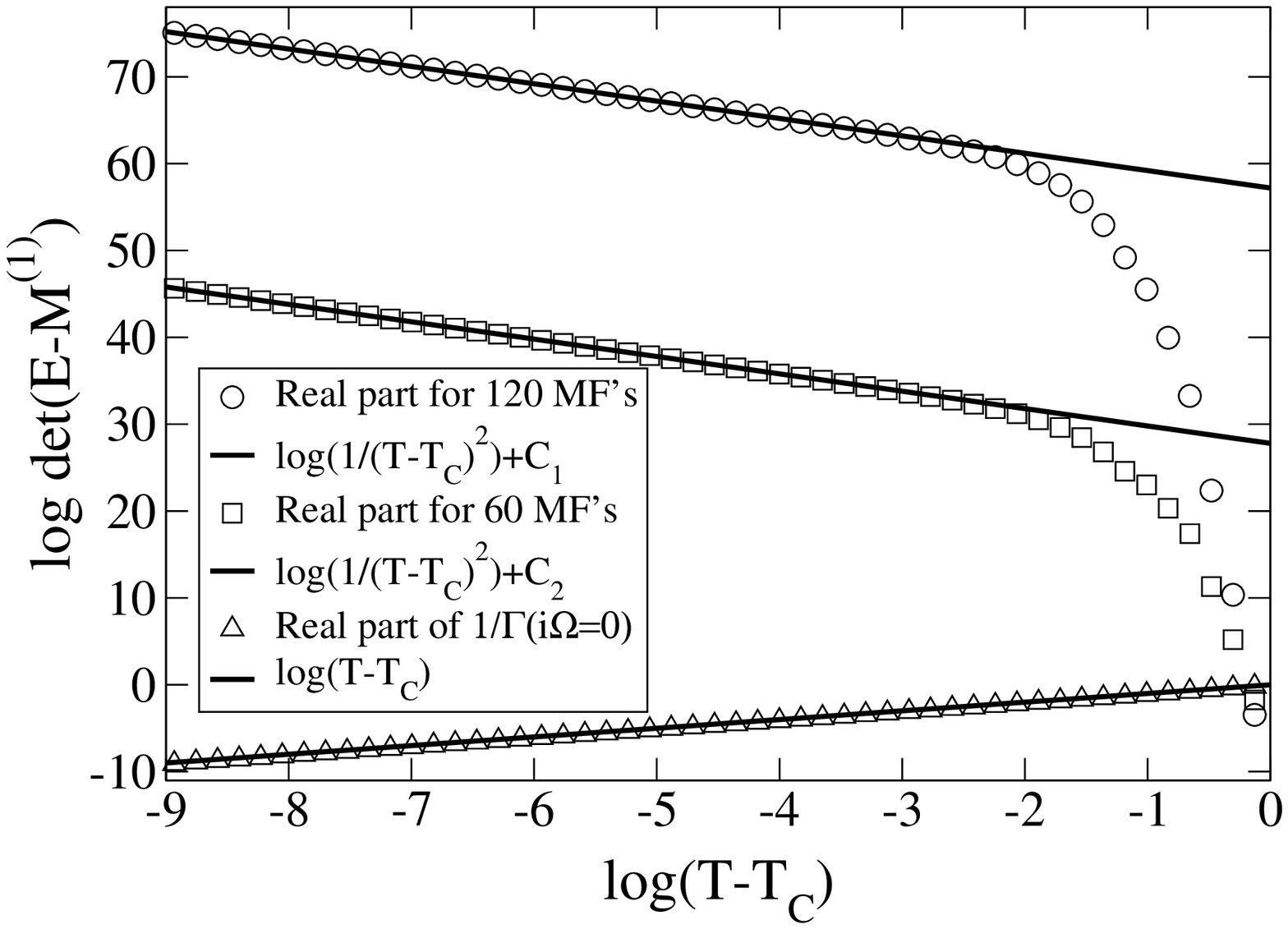}}}}
\caption{The determinant given in eq. (\protect{\ref{eq:det}})
diverges at the Thouless instability proprtional to
$(T-T_c)^{-2}$. This is shown for $\tilde{\mu} = -0.5$ and $T_c
\approx 0.37$ for
two maximum numbers of Matsubara
frequencies $M=60$ and $M=120$. The total frequency $i \omega_n^g$ was
choosen at $n=1$. The attractive interaction $U$ is equal to the bare
bandwidth $W$. For a comparison the value of $1+U\;\chi$ which goes to
zero linearly in T, with $(T-T_c)$, is plotted.
}
\label{fig:det}
\end{figure}

\end{document}